\documentclass[reprint,aps,pra,superscriptaddress,nofootinbib,preprintnumbers,longbibliography]{revtex4-1}

\usepackage{comment}
\usepackage{graphics}
\usepackage{bm}
\usepackage{color}
\usepackage{amsmath}
\usepackage{amssymb}
\usepackage{mathrsfs}
\usepackage{graphicx}
\usepackage[caption=false]{subfig}
\usepackage{enumitem}
\usepackage{hyperref} 
\usepackage{multirow}

\begin{document}
\title{Single Electrons in a Dual-Plane Printed-Circuit-Board Penning Trap}

\date{\today}

\author{Zirui Fang}
\affiliation{Department of Physics, Harvard University, Cambridge, Massachusetts 02138, USA}

\author{Benedict A. D. Sukra}
  \affiliation{Center for Fundamental Physics, Department of Physics and Astronomy, Northwestern University, Evanston, Illinois 60208, USA}
 
\author{Xing Fan}
\email{xingfan@g.harvard.edu}
\affiliation{Department of Physics, Harvard University, Cambridge, Massachusetts 02138, USA}

\begin{abstract}
We demonstrate single-electron trapping and detection in a two-dimensionally scalable dual-plane printed-circuit-board Penning trap. We characterize deterministic electron loading, axial damping, axial temperature, and collision-induced magnetron-radius growth at low magnetic fields. These results establish a practical platform for planar Penning traps and identify key next steps toward applications in quantum information science.
\end{abstract}
\maketitle
%introduction%
\section{Introduction}
Isolated electrons in a Penning trap are a good system for precision measurement \cite{1987DehmeltMagneticMoment,RevModPhys.58.233,ElectronMagneticMoment_Fan_PRL_2023,odom2006new,hanneke2008new,QLS_Electron_2025,Atoms_Review_2019_g-factor}, dark matter searches\cite{Fan_DarkPhoton_Axion_DarkMatter_2022,HighlyExcitedDPAxion_Fan2025,Imperial_ElectronDarkPhoton_2026}, and quantum computing \cite{Feasibility_PRA_Haeffner_2022, GoldmanPRAPRoposal2010, Goldman2011-PlanarTrapQIS,HaeffnerElectronTrapping2021}.
Since the motional coupling rate between two identical ions via image charges is inversely proportional to their mass \cite{sorensen2004capacitive,QLS_Electron_2025,QLS_PhysRevA_1990}, trapped electrons are an attractive quantum information platform---being 80,000 times lighter than Ca$^+$ and 300,000 times lighter than Yb$^+$ \cite{zurita2008wiring,osada2022feasibility,GoldmanPRAPRoposal2010,Mancini_PhysRevA_1999,Ciaramicoli_PhysRevA_2001,Ciaramicoli_JModOpt_2002,Ciaramicoli_PhysRevLett_2003,Ciaramicoli_PhysRevA_2004,Ciaramicoli_PhysRevA_2005,Stahl_EurPhysJD_2005,Galve_EurPhysJD_2006,Bushev_EurPhysJD_2008}. Consequently, quantum gates using electrons could be much faster than those using conventional atomic ions.
Due to their all-DC trapping principle, Penning traps also provide a good environment for quantum control \cite{HeatingRateReview2015}.
Unlike RF ion traps, Penning traps have no micromotion, and low heating rates have been measured in cryogenic Penning traps \cite{RevModPhys.87.1419,PhysRevLett.122.043201,PhysRevResearch.6.033233,JonathanHome_SurfaceHeatingRate_Nature_2024,JonathanHome_3DScan_2024}.
The operation of Penning traps at cryogenic temperatures has been well established, further suppressing the heating rate \cite{tan1989one,alphatrap,smorra2015base,latacz2023base,PhysRevLett.124.113001,fan2022_Thesis,smorra2015base,nagahama2017sixfold,smorra2017spintransitions,smorra2017ppb,borchert2022chargeMass,leonhardt2025transport,latacz2025coherent}.

However, electrons in Penning traps have mainly been used for tests of fundamental physics, such as $g$-factor measurements \cite{1987DehmeltMagneticMoment,odom2006new,hanneke2008new,ElectronMagneticMoment_Fan_PRL_2023} and the search for physics beyond the Standard Model including dark-photons and axions \cite{Fan_DarkPhoton_Axion_DarkMatter_2022,HighlyExcitedDPAxion_Fan2025,Imperial_ElectronDarkPhoton_2026,Carney2021_TrappedIonDetector}.
The application to quantum information science (QIS) has been discussed in Refs.~\cite{zurita2008wiring,osada2022feasibility,QLS_Electron_2025,PhysRevApplied.22.024032,daniilidis2013quantum}, but it has not, to our knowledge, been realized.
Penning traps for fundamental physics require large three-dimensional geometries to achieve high sensitivity \cite{RevModPhys.58.233,Fang_sphericalTrap_2026,gabrielse1984cylindrical,brown1985geonium,hanneke2011cavity,alphatrap,latacz2023base}, which is difficult to scale for QIS applications.
In fact, scalable 2D surface Penning traps have been proposed and discussed in Refs.~\cite{daniilidis2009wiring,zurita2008wiring,osada2022feasibility,PhysRevApplied.22.024032}, but single-electron detection has never been achieved due to their large electric field anharmonicity.

In this paper, we report single-electron detection and characterization in a dual-plane surface Penning trap.
The trap consists of two printed circuit boards (PCBs) placed mirror-symmetrically to achieve a highly harmonic trap potential (Sec.~\ref{Sec:Detection}).
The realized high electric-field harmonicity reduces axial frequency fluctuation and allows single electron detection \cite{GoldmanPRAPRoposal2010,Bushev_EurPhysJD_2008}.
The PCB design allows integration of other electronics on the rear side, advantageous for the control of future 2D multiplexed traps.
Additionally, we characterize the magnetron motion of single electrons at extremely low magnetic fields, $B<0.2$~T, and observe collision-induced growth of the magnetron orbit (Sec.~\ref{Sec:Magnetron}).
We discuss the next steps in Sec.~\ref{Sec:Discussion} and summarize this paper in Sec.~\ref{Sec:Summary}.

\section{Trap Design and Single Electron Detection}
\label{Sec:Detection}
\begin{figure*}
    \centering
\includegraphics[width=0.95\linewidth]{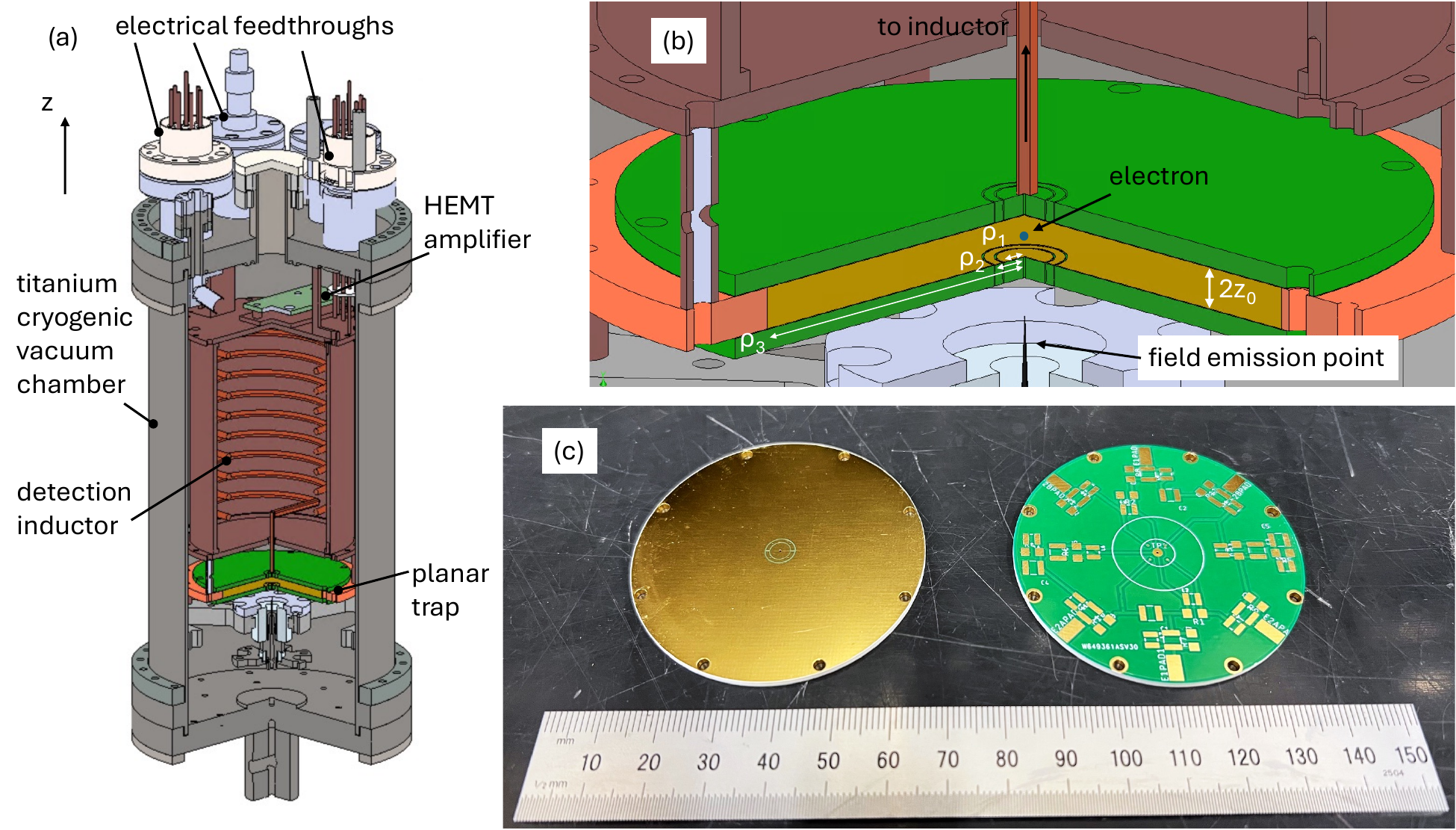} 
     \caption{(a) Cryogenic section of the dual-plane Penning trap, (b) zoom-in near the trap assembly , and (c) the actual PCBs. The electrodes are segmented into three pads, e1, e2, and e3 from the center, with radii $\rho_1$, $\rho_2$, and $\rho_3$ respectively, and are placed mirror-symmetrically with a copper ring spacer.}
    \label{fig:Design}
\end{figure*}

Fig.~\ref{fig:Design} shows the design of the planar trap (a,b) and the fabricated PCBs (c).
Two PCBs are placed symmetrically on the top and bottom of the copper spacer ring.
Each plane is divided into three electrode rings, labeled e1, e2, and e3 from the center with radii $\rho_1$, $\rho_2$, and $\rho_3$, respectively.
The PCBs are 1.5-mm-thick, gold-plated Rogers 4003C boards with a 0.3~mm central hole for efficient electron loading by a field emission point (FEP) placed below the trap.
The electrodes are biased from the rear side through 0.2-mm vias.
Additionally, the e2 electrodes are split into two to enable efficient application of an axial-magnetron coupling drive\cite{RevModPhys.58.233}.

A detection inductor ($l=600$~nH) is connected to the top e1 electrode.
Together with the parasitic capacitance of the electrode ($c\sim$6~pF), they form a resonator with a quality factor of $Q=1450$ at the axial frequency of the trapped particle $\omega_z/2\pi=81.6$~MHz.
The resonator is coupled to a high-electron-mobility transistor (HEMT) amplifier (FHX13LG, Fujitsu)\cite{durso2003Thesis}.
The bias power for the HEMT amplifier used in this work is $P=24$~$\mu$W unless otherwise specified.
The entire assembly [Fig. \ref{fig:Design}(a)] of the planar trap, FEP, and the resonator is housed in a titanium cryogenic vacuum chamber and cooled to 4.5~K using a pulse-tube cryocooler (JMTE-Insert, JASTEC, Inc.).
A sweepable homogeneous magnetic field ($B=0$--6~T) is applied in the $z$-direction using a superconducting magnet (JMTD-6T152SS, JASTEC, Inc.).

The radii of the electrodes and the interplanar distance $2z_0$ determine the electrostatic potential~\cite{gabrielse1984cylindrical,PhysRevA.27.2277,GoldmanPRAPRoposal2010}.
Using $r=\sqrt{x^2+y^2+z^2}$, $\rho=\sqrt{x^2+y^2}$, and $\theta=\cos^{-1}(z/r)$, the potential near the center is written as
% \begin{subequations}
\begin{align}
\phi(\rho,z)=&-V_1\sum_{\substack{k=0}}C_{k}\left(\frac{r}{z_0}\right)^kP_k(\cos\theta)\nonumber\\
=&-V_1\sum_{\substack{k=0}}\left(C^0_k+D_k\frac{V_2}{V_1}\right)\left(\frac{r}{z_0}\right)^kP_k(\cos\theta),
\end{align}
% \end{subequations}
where $P_k(x)$ is the Legendre polynomial function, $V_1$ and $V_2$ are the applied voltages to e1 and e2 respectively with e3 and the copper spacer grounded, and $C^0_k$ and $D_k$ are geometrical parameters reflecting the $V_1$ and $V_2$ biases.
$C_2$ is the harmonic potential term and $C_i$ ($i\geq3$...) are the anharmonicity terms.
$C_2$ determines the axial frequency by
\begin{equation}
    \omega_z=2\pi\nu_z=\sqrt{\frac{2C_2eV_1}{m_ez_0^2}},
\end{equation}
where $e$ is the elementary charge and $m_e$ is the electron's mass.
When the cyclotron orbit and the magnetron orbit are much smaller than the axial oscillation amplitude, the electric potential can be simplified along the $z$-axis,
\begin{equation}
\phi(z)=-V_1\sum_{\substack{k=0}}C_{k}\left(\frac{z}{z_0}\right)^k\\
\end{equation}
With the anharmonicity, the axial frequency depends on the axial oscillation amplitude $z_A$ as \cite{gabrielse1984cylindrical}
\begin{eqnarray}    
\Delta\nu_z&=&\nu_z\Bigg[\left(-\frac{15C_3^2}{16C_2^2}+\frac{3C_4}{4C_2}\right)\left(\frac{z_A}{z_0}\right)^2\nonumber\\
&+&\left(-\frac{15C_3^3}{16C_2^3}+\frac{3C_3C_4}{4C_2^2}\right)\left(\frac{z_A}{z_0}\right)^3\nonumber\\
&+&\Bigg(-\frac{2565}{1024}\frac{C_3^4}{C_2^4}
+\frac{645}{128}\frac{C_3^2C_4}{C_2^3}
-\frac{105}{32}\frac{C_3C_5}{C_2^2}\nonumber\\
&&-\frac{21}{64}\frac{C_4^2}{C_2^2}
+\frac{15}{16}\frac{C_6}{C_2}
\Bigg)
\left(\frac{z_A}{z_0}\right)^4+\cdots\Bigg]
\end{eqnarray}

In previous single-plane geometries, due to the lack of symmetry, the leading-order anharmonicity $C_3$ coupled to the probability distribution of the axial oscillation amplitude from the Boltzmann distribution 
\begin{eqnarray}    
    p(z_A^2)&=&\frac{m_e\omega_z^2}{2k_BT_z}\exp\left(-\frac{m_e\omega_z^2z_A^2}{2k_BT_z}\right)\label{eq:pz_A2}\\
\langle z_A^2\rangle&=&\frac{2k_BT_z}{m_e\omega_z^2}
\end{eqnarray}
and generated axial frequency fluctuation \cite{goldman2011Thesis,MelissaWessels2019Thesis}.
Even with the well-designed three-gap single-plane geometry \cite{GoldmanPRAPRoposal2010}, imperfections of the electrodes and the surrounding shield create uncontrolled anharmonicity \cite{MelissaWessels2019Thesis}.
Instead, in the dual-plane configuration, the mirror symmetry makes all odd terms ($C_3$, $C_5$, ...) negligibly small, allowing intrinsically more harmonic trapping potentials.
The leading term of the frequency shift simplifies to
\begin{equation}    
    \Delta\nu_z=\nu_z\frac{3C_4}{4C_2}\frac{z_A^2}{z_0^2}\label{eq:DeltaNu_z_Shift}.
\end{equation}
Thus, the averaged axial frequency broadening is
\begin{equation}
    \langle\Delta\nu_z\rangle=\nu_z\frac{3C_4}{2C_2}\frac{k_BT_z}{m_e\omega_z^2z_0^2}.
\end{equation}
In the dual-plane geometry, $C^0_k$ and $D_k$ are given analytically by \cite{GoldmanPRAPRoposal2010}
\begin{subequations}
    \begin{align}
    C_k^0=&\frac{-2}{k!}\left(\frac{z_0}{\rho_3}\right)^{k}
    \sum_{n=1}^\infty\frac{x_{0n}^{k-1}}{\rho_3\left[J_1(x_{0,n})\right]^2\cosh\!\left(\frac{x_{0n}}{\rho_3}z_0\right)}\nonumber\\
    &\times\rho_1J_1\!\left(\frac{x_{0n}\rho_1}{\rho_3}\right).\\
    D_k=&\frac{-2}{k!}\left(\frac{z_0}{\rho_3}\right)^{k}
    \sum_{n=1}^\infty\frac{x_{0n}^{k-1}}{\rho_3\left[J_1(x_{0,n})\right]^2\cosh\!\left(\frac{x_{0n}}{\rho_3}z_0\right)}\nonumber\\
    &\times\left[\rho_2J_1\!\left(\frac{x_{0n}\rho_2}{\rho_3}\right)-\rho_1J_1\!\left(\frac{x_{0n}\rho_1}{\rho_3}\right)\right].
    \end{align}
\end{subequations}
We design the e2 electrodes to be $D_2=0$ and $D_4
\neq0$ (the so-called orthogonality condition), allowing tuning of the leading anharmonicity $C_4$ without changing $C_2$ and thus $\omega_z$ \cite{gabrielse1984cylindrical,tan1989one}.
When $C_4$ is tuned to 0 by setting $V_2=-(C^0_4/D_4)V_1$, the axial frequency broadening is minimum, allowing the resolution of single-electron signals. The anharmonicity parameters for our trap are summarized in Tab.~\ref{tab:anharmonicity}.
Experimentally, we found that $V_1=-12.48$~V and $V_2=-16.15$~V give $\omega_z/(2\pi)=81.6$~MHz and $C_4/C_2<10^{-3}$.
The slight deviation from the calculation is due to the mechanical tolerance of the copper spacer, the gaps on the PCBs, and thermal contraction.
\begin{table}
\caption{Calculated anharmonicity parameters for the described dual-plane geometry with $\rho_1=2.00$~mm, $\rho_2=2.85$~mm, $\rho_3=25.40$~mm, and $z_0=3.00$~mm. The right-hand columns show the coefficients when $V_2$ is tuned to achieve the harmonic potential ($C_4=0$).}
\label{tab:anharmonicity}
\centering
\begin{tabular}{|c|c||c|c|}
\hline
parameter & value & parameter&value when $V_2=-(C^0_4/D_4)V_1$\\
\hline
$C_2^0$     & $-0.519$     &\multirow{2}{*}{$C_2$}   &\multirow{2}{*}{$-0.519$}\\
$D_2$       & 0.000   &   &\\
\hline
$C_4^0$     & $-0.281$    &\multirow{2}{*}{$C_4$}   &\multirow{2}{*}{0}\\
$D_4$       & $0.2037$    &   &\\
\hline
$C_6^0$     & $-0.003$    &\multirow{2}{*}{$C_6$}   &\multirow{2}{*}{$0.101$}\\
$D_6$       & 0.076     &   &\\
\hline
$C_8^0$     & 0.091     &\multirow{2}{*}{$C_8$}   &\multirow{2}{*}{$-0.007$}\\
$D_8$       & $-0.071$     &   &\\
\hline
\end{tabular}

\end{table}

\begin{figure}[t]
    \centering
\includegraphics[width=0.8\linewidth]{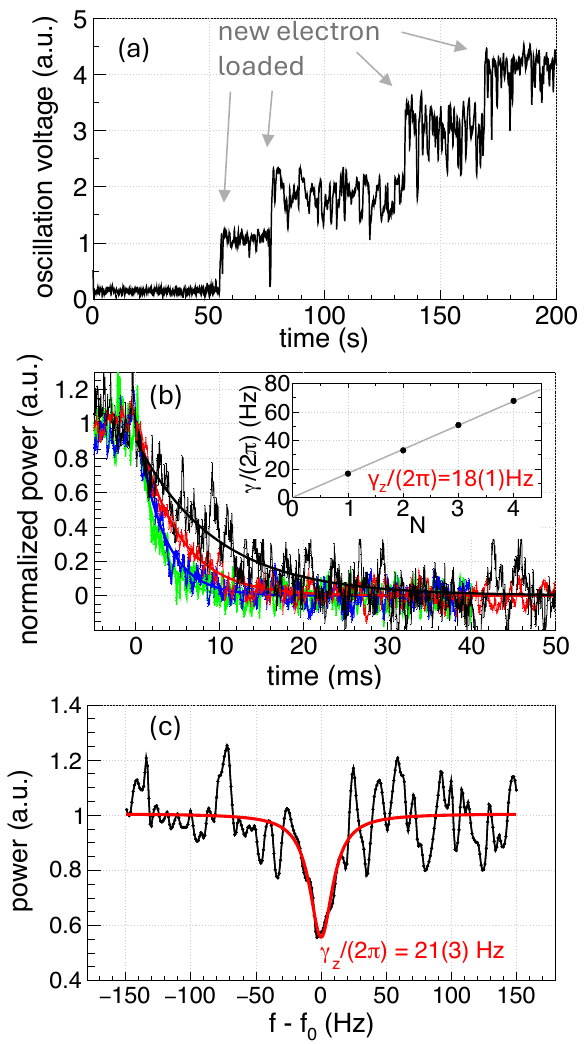} 
     \caption{(a) Deterministic loading of single electrons. A strong parametric drive at $\omega_\text{d}=2\omega_z$ is applied while the FEP is weakly and continuously fired. (b) Measurement of $\gamma=N\gamma_z$ for $N=1$~(black), 2~(red), 3~(blue), and 4~(green) electrons. The inset shows the fitted $\gamma$ as a function of $N$. (c) Fourier transform of the axial amplifier output showing a single electron's dip. The measured $\gamma_z$ is consistent with (b) but slightly larger due to residual axial frequency fluctuation due to the pulse-tube cryocooler.}
    \label{fig:SingleElectron}
\end{figure}

Fig.~\ref{fig:SingleElectron} shows the signals from single electrons at $B=5.06$~T.
The magnetic field is chosen to minimize the potential magnetron heating, as discussed in Sec.~\ref{Sec:Magnetron} and to achieve fast synchrotron radiation cooling for the cyclotron  \cite{brown1985cyclotron,brown1985cyclotronPRL,gabrielse1985observation}.
In Fig.~\ref{fig:SingleElectron}(a), a strong parametric drive at $\omega=2\omega_z$ is applied, and the FEP is weakly and continuously fired.
Each step-increase of the amplitude corresponds to the arrival of a new electron.
By monitoring the step-increase, we can load any desired number of electrons.
For $N\geq2$, the parametrically driven response is noisier due to the internal motion of the trapped particles. This internal motion is mainly cooled by synchrotron radiation at a rate determined by the trap's microwave cavity  \cite{brown1985cyclotron,brown1985cyclotronPRL,brown1986cyclotronSpherical,Dehmelt_MicrowaveCavity_Obseration_1987}.
For our planar trap in its current configuration, due to a weaker microwave resonant structure, the cooling rate is lower than a cylindrical trap \cite{tan1989one} or a spherical trap \cite{Fang_sphericalTrap_2026}, resulting in less stable parametric response.
Nonetheless, clear deterministic counting is easily possible from Fig.~\ref{fig:SingleElectron}(a).

In Fig.~\ref{fig:SingleElectron}(b), after the deterministic loading of $N$ electrons, we measure its axial damping constant $\gamma=N\gamma_z$. Hereafter, we use $\gamma_z$ for the single-electron damping constant and $\gamma$ for the general $N$ electrons.
To measure $\gamma$, the axial oscillation is excited by a strong parametric drive to about $z_A=100(15)$~$\mu$m, and the drive is turned off at $t=0$~ms.
The excited axial motion damps its oscillation energy to the detection circuit with a rate $\gamma$.
This method has an advantage that the measured damping rate is insensitive to axial frequency fluctuation smaller than the detection bandwidth ($\sim$16~kHz).
Each color in Fig.~\ref{fig:SingleElectron}(b) corresponds to a different number of electrons, and one can see discrete increases of $\gamma$ (inset), giving $\gamma_z/(2\pi)=18 (1)$~Hz.

The single electron damping rate is given by \cite{ElectronCalorimeter}
\begin{equation}
    \gamma_z=\frac{1}{m_e}\left(\frac{ec_1}{2z_0}\right)^2R,\label{eq:gamma_z}
\end{equation}
where $R=Q\omega_zl=450$~k$\Omega$ is the effective parallel resistance of the detection resonator and $c_1=0.56$ is the image charge pickup constant.
We estimate $\gamma_z/(2\pi)=18(4)$~Hz, where the uncertainty comes from the estimate of the parasitic capacitance $c$.
For $N$ trapped electrons, the damping rate is given by $\gamma=N\gamma_z$, as visible from Fig.~\ref{fig:SingleElectron}(b).

Fig.~\ref{fig:SingleElectron}(c) shows the dip of a single electron without any excitation drive.
The electron in this setting is thermalized to the resonant circuit, creating a dip with a width $\gamma_z$ at its axial frequency \cite{ElectronCalorimeter}. The measured $\gamma_z/(2\pi)=21 (3)$~Hz in Fig.~\ref{fig:SingleElectron}(c) gives slightly larger $\gamma_z$ than that in Fig.~\ref{fig:SingleElectron}(b) due to fluctuations of the axial frequency caused by the pulse-tube cryocooler.
Still, this measurement confirms the measured $\gamma_z$ in Fig.~\ref{fig:SingleElectron}(b) and the capability of single-electron detection.

\begin{figure}[t]
    \centering
\includegraphics[width=0.85\linewidth]{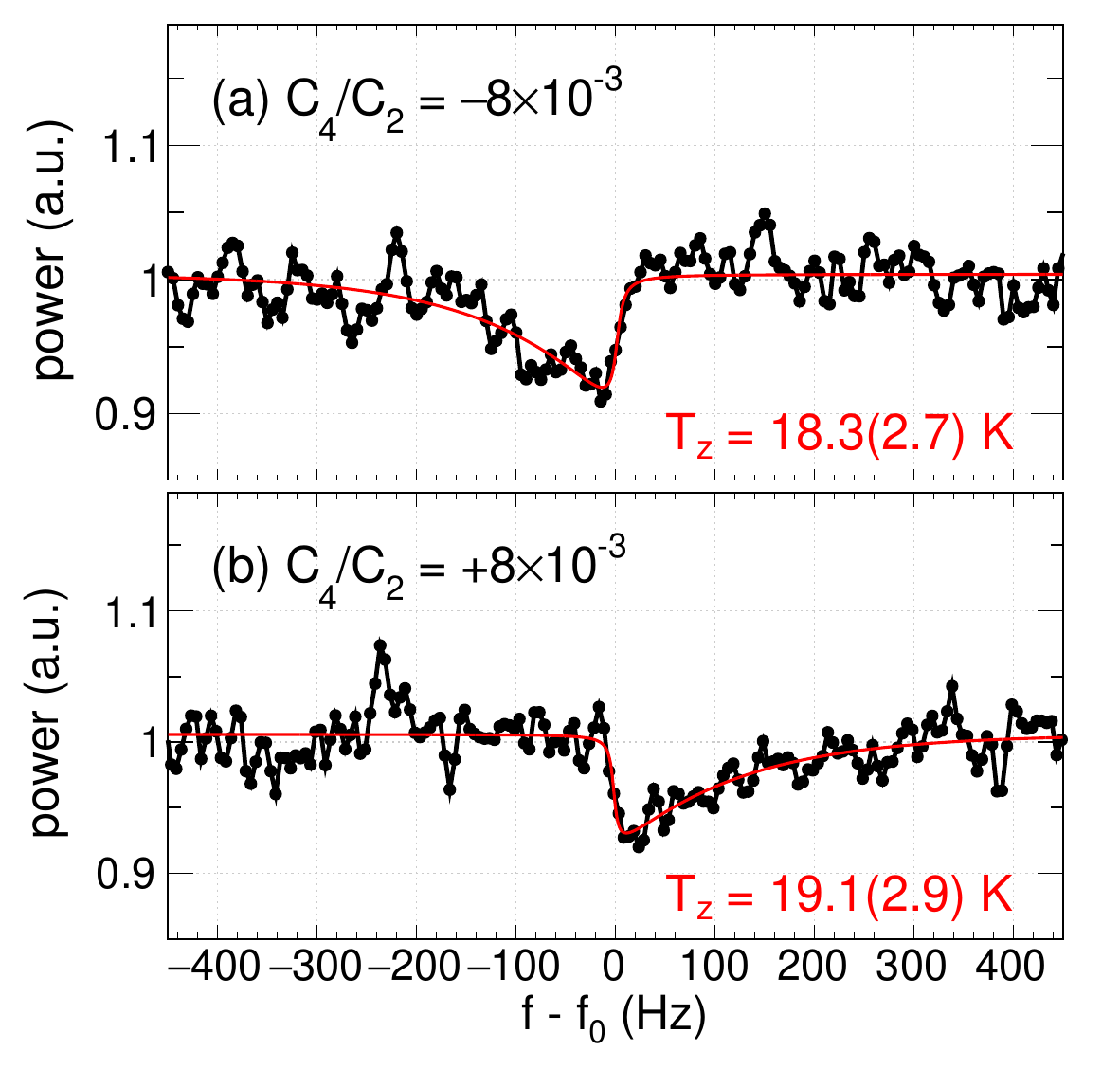} 
     \caption{Measurement of the axial temperature $T_z$ from the dip shape and fitting by Eq.~\eqref{eq:Convolution} for (a) $C_4/C_2=-8\times10^{-3}$ ($\Delta V_2=+0.25$~V) and (b) $C_4/C_2=+8\times10^{-3}$ ($\Delta V_2=-0.25$~V).}
    \label{fig:Temperature}
\end{figure}
Additionally, the electron's axial temperature $T_z$ can be determined by deliberately introducing anharmonicity $|C_4|>0$ through changing $V_2$.
Since the electron is strongly thermalized to the detection circuit, $T_z$ is determined by the physical temperature as well as the equivalent input temperature of the HEMT amplifier.
In thermal equilibrium, the Boltzmann distribution of the axial oscillation amplitude [Eq.~\eqref{eq:pz_A2})] couples to the axial frequency shift [Eq.~\eqref{eq:DeltaNu_z_Shift})], changing the dip shape to a convolution of a Lorentzian and a Boltzmann distribution,
\begin{equation}
    f(\nu)=\int d\nu'\,L(\nu',\nu_z,\gamma_z)B(\nu-\nu',\langle\Delta\nu_z\rangle),\label{eq:Convolution}
\end{equation}
where
\begin{eqnarray}    
    L(\nu,\nu_z,\gamma_z)&=&\frac{1}{\pi}\frac{\frac{1}{2}\frac{\gamma_z}{2\pi}}{\left(\nu-\nu_z\right)^2+\left(\frac{1}{2}\frac{\gamma_z}{2\pi}\right)^2}\\
    B(\nu,\langle\Delta\nu_z\rangle)&=& \frac{1}{|\langle\Delta\nu_z\rangle|}\exp\left(-\frac{\nu}{\langle\Delta\nu_z\rangle}\right)\nonumber\\
    &\times&\begin{cases}
       H(\nu)~~~~~\,\text{for}~~\langle\Delta\nu_z\rangle>0\\
        H(-\nu)~~~\text{for}~~\langle\Delta\nu_z\rangle<0\\
    \end{cases},
\end{eqnarray}
and $H(x)$ is the Heaviside step function.

Fig.~\ref{fig:Temperature} shows the measured single-electron dip for intentionally detuned $C_4$ values and fitting curves using Eq.~\eqref{eq:Convolution}.
One can indeed see the asymmetric Boltzmann distribution.
The fitted temperature $T_z\approx18$~K is much higher than the physical temperature.
This has also been observed in Refs.~\cite{odom2004Thesis,durso2003Thesis}, and is likely due to the strong coupling to the HEMT amplifier's input noise.
The power dissipation of the HEMT amplifier is $P=24$~$\mu$W, which is similar to the conditions in Refs.~\cite{odom2004Thesis,durso2003Thesis}.
We also note that the trapping potential must be detuned as much as $\Delta V_2=40$~mV to generate a shift of $|\langle\Delta\nu_z\rangle|=18$~Hz to broaden the axial dip, demonstrating the relaxed requirements for $V_2$ due to the mirror-symmetric design.

\section{Collision-Induced Magnetron Radius Growth}
\label{Sec:Magnetron}
With proposed QIS applications at low cyclotron frequencies and $N\geq2$ electrons in mind \cite{Home_ScalableArray_PRX_2020,lamata2010towards}, we report an observation of excessive growth of the magnetron radius at low $B$-fields and for $N\geq2$ electrons.
We study the dependence on the number of trapped electrons $N$, magnetic field $B$, the amplifier bias power $P$, and the axial oscillation amplitude, and interpret the growth as collision-induced.
While this effect may not be specific to the planar geometry, the study in the planar geometry should be directly relevant for the proposed applications in QIS \cite{zurita2008wiring,osada2022feasibility,QLS_Electron_2025,PhysRevApplied.22.024032,GoldmanPRAPRoposal2010,Mancini_PhysRevA_1999,Ciaramicoli_PhysRevA_2001,Ciaramicoli_JModOpt_2002,Ciaramicoli_PhysRevLett_2003,Ciaramicoli_PhysRevA_2004,Ciaramicoli_PhysRevA_2005,Stahl_EurPhysJD_2005,Galve_EurPhysJD_2006,Bushev_EurPhysJD_2008}.

\begin{figure}[t]
    \centering
\includegraphics[width=0.95\linewidth]{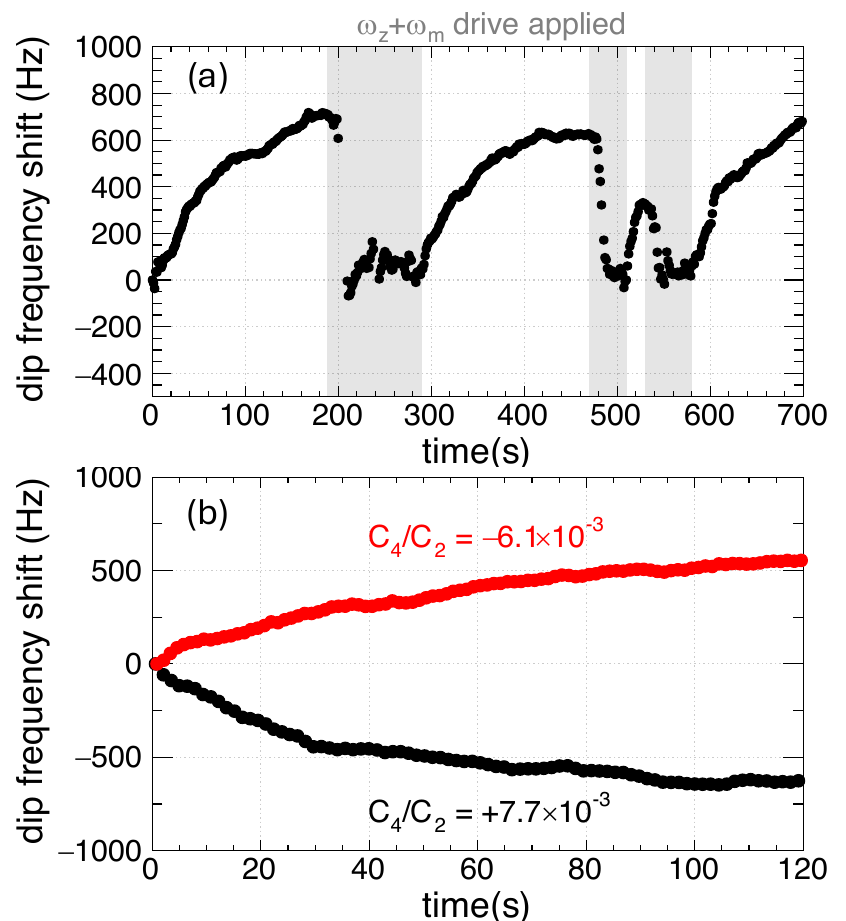} 
     \caption{(a) Monitored axial dip frequency as a function of time with $C_4/C_2=-6.1\times10^{-3}$. Axial-magnetron coupling drives at $\omega_z+\omega_m$ are applied in the shaded region to reduce magnetron radius to the minimum. (b) Shift of the dip frequency for $C_4/C_2=-6.1\times10^{-3}$ and $C_4/C_2=+7.7\times10^{-3}$, showing the reversal of the drift direction when flipping $C_4$. Other conditions in these measurements are $N=7$, $B=0.045$~T, and $P=24$~$\mu$W.}
    \label{fig:MagnetronHeating_DriveAndC4}
\end{figure}
In an anharmonic trap, the axial frequency depends not only on the axial amplitude but also on the orbital radius\cite{RevModPhys.58.233,durso2003Thesis,goldman2011Thesis}.
For example, the leading term $C_4$ leads to
\begin{eqnarray}    
    \phi(\rho,z)=&-&V_1C_2\left(\frac{z^2-\rho^2/2}{z_0^2}\right)\nonumber\\
    &-&V_1C_4\left(\frac{z^4-3z^2\rho^2+3\rho^4/8}{z_0^4}\right).
\end{eqnarray}

Here, we are interested in the effect of a large orbit $\rho$, so we assume $\rho\gg z$, resulting in a $\rho$-dependent axial frequency shift
\begin{equation}
    \Delta\nu_z=-\nu_z\frac{3}{2}\frac{C_4}{C_2}\frac{\rho^2}{z_0^2}.
\end{equation}
With a typical $C_4/C_2=-6.1\times10^{-3}$ in this section, $\rho_m=80~\mu$m corresponds to $\Delta\nu_z=500$~Hz shift.

The magnetron orbit can be reduced by applying an axial-magnetron coupling drive at $\omega_z+\omega_m$, where $\omega_m$ is the magnetron frequency \cite{RevModPhys.58.233}.
After this drive, the magnetron radius is $\rho_m=\sqrt{\frac{\omega_m}{\omega_z}\frac{4k_BT_z}{m_e\omega_z^2}}$.
Even with $B=0.034$~T (the lowest magnetic field in the following) and $T_z=18$~K, the radius is $\rho_m=13~\mu$m and the shift is only $\Delta\nu_z=15$~Hz, much smaller than any shifts we observe here.

Fig.~\ref{fig:MagnetronHeating_DriveAndC4} shows the monitored dip frequency for $N=7$, $B=0.045$~T, and $P=24$~$\mu$W.
In Fig.~\ref{fig:MagnetronHeating_DriveAndC4}(a), $C_4/C_2=-6.1\times10^{-3}$, and repeated axial-magnetron drives at $\omega_z+\omega_m$ are applied in the shaded region \cite{RevModPhys.58.233}.
During each drive, the axial frequency is restored to the same value, and drifts again at the same rate.
In Fig.~\ref{fig:MagnetronHeating_DriveAndC4}(b), we monitor the drift with the opposite sign of $C_4$, confirming that the shift of the axial dip frequency indeed reverses with $C_4$.
Based on these two observations, we infer that the axial frequency shift is induced by an increase in the magnetron orbital radius.

\begin{figure}[t]
    \centering
\includegraphics[width=0.95\linewidth]{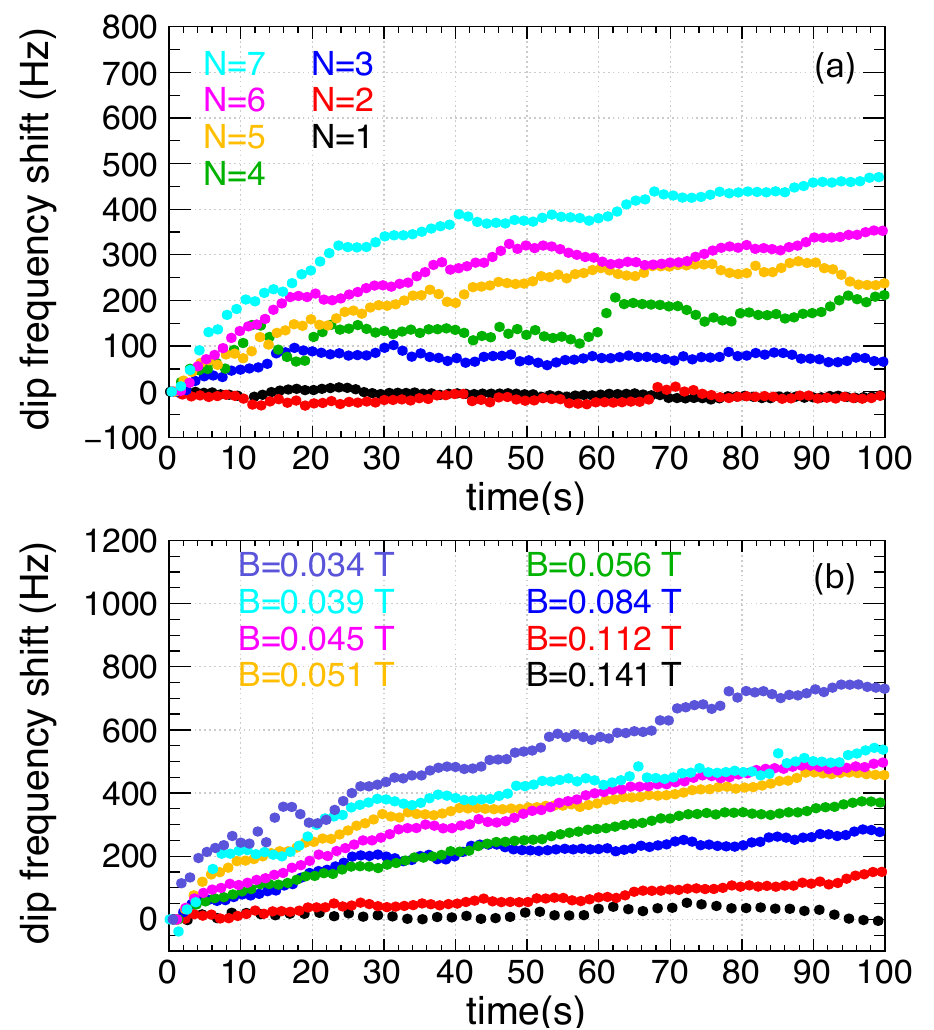} 
     \caption{(a) Monitored axial dip frequency as a function of time for different numbers of electrons $N$ with $B=0.045$~T, $C_4/C_2=-6.1\times10^{-3}$, and $P=24$~$\mu$W. (b) The same measurement for different magnetic fields with $N=7$, $C_4/C_2=-6.1\times10^{-3}$, and $P=24$~$\mu$W.}
    \label{fig:MagnetronHeating_NAndB}
\end{figure}
The observed axial frequency shift depends on the number of trapped electrons $N$ [Fig.~\ref{fig:MagnetronHeating_NAndB}(a)].
In these measurements, the axial-magnetron coupling drive is turned off at $t=0$~s.
The monitored axial dip frequency is considerably more stable for fewer electrons.
Based on this measurement, we infer that the observed magnetron radius growth is induced by the collisions among the trapped electrons.

The axial frequency drift also depends on the magnetic field.
In Fig.~\ref{fig:MagnetronHeating_NAndB}(b), the magnetic field is lowered gradually.
Within the measurement time window (100~s), as the magnetic field is lowered below 0.1~T, the axial frequency shift is observable.
Below 0.034~T, the axial frequency drift is too fast, and we were not able to resolve a clear dip in a short enough averaging time.

\begin{figure}[t]
    \centering
\includegraphics[width=0.95\linewidth]{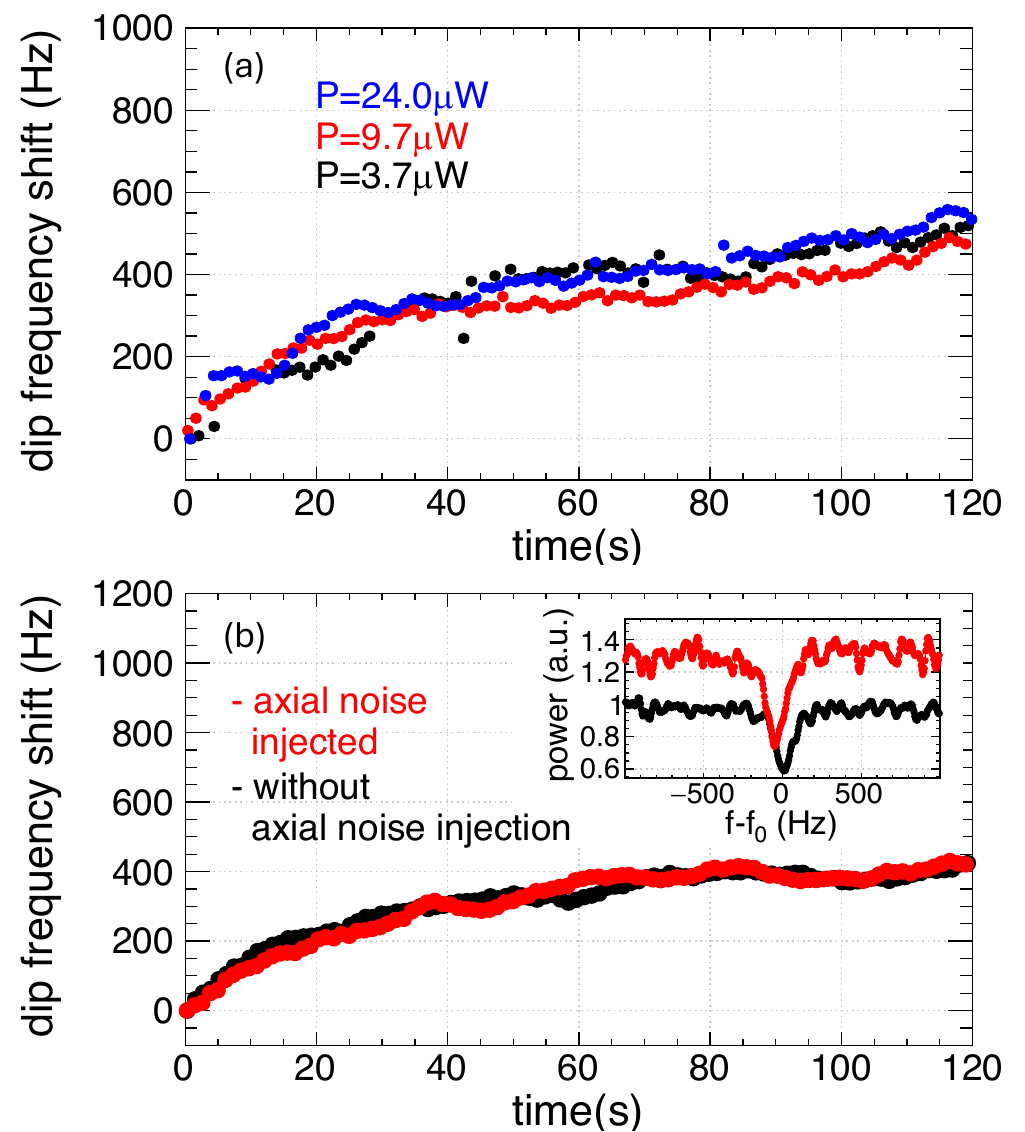} 
     \caption{(a) Monitored axial dip frequency as a function of time for different amplifier bias power $P$ with $B=0.045$~T, $C_4/C_2=-6.1\times10^{-3}$, and $N=7$. (b) The same measurement with intentionally injected noise at the axial frequency $\nu_z\pm8$~kHz. The dip frequency shift by the axially injected drive (visible from inset) is subtracted so that both drift measurements start from 0~Hz at $t=0$~s. $N=7$, $B=0.045$~T, $C_4/C_2=-6.1\times10^{-3}$, and $P=24$~$\mu$W.}
    \label{fig:MagnetronHeating_P_Noise}
\end{figure}
To determine whether the collision-induced effect originates from the axial motion or from the cyclotron motion, we monitor the axial dip frequency with different amplifier bias powers $P$ [Fig.~\ref{fig:MagnetronHeating_P_Noise}(a)] and with intentionally injected white noise around $\nu_z$ [Fig.~\ref{fig:MagnetronHeating_P_Noise}(b)].
In Fig.~\ref{fig:MagnetronHeating_P_Noise}(a), the first stage amplifier bias is set differently for each curve.
Due to the worse signal-to-noise ratio at lower $P$, we are not able to extract $T_z$ with lower bias powers, but it is known that changing the amplifier bias changes the axial motion's temperature $T_z$ \cite{odom2006new,durso2003Thesis,hanneke2007Thesis}.
No change in the growth rate of the magnetron radius is observed.
In Fig.~\ref{fig:MagnetronHeating_P_Noise}(b), we intentionally inject additional white noise in the range of $\nu_z\pm8$~kHz using a white noise generator and a crystal filter.
The inset in Fig.~\ref{fig:MagnetronHeating_P_Noise}(b) shows the Fourier transform spectrum, and one can indeed see a higher noise level and shift of the dip frequency due to axial excitation.
The excited axial amplitude is estimated to be $z_A=120~\mu$m from the shift of the axial dip frequency, 2.5 times higher than the average axial amplitude in the Boltzmann distribution $\sqrt{2k_BT_z/(m_e\omega_z^2)}=47$~$\mu$m.
Again, in this comparison, no change in the magnetron growth rate is observed.
Based on these two measurements, we believe that the collision-induced axial frequency shift does not originate from the axial motion temperature.

Currently, we are not able to check the dependence on the cyclotron motion.
The cyclotron frequency at these magnetic fields is 0.8--3~GHz.
The gap between the two planes is 6~mm, which makes it difficult for the cyclotron drive to reach the trapped electrons.
The excited cyclotron motion also couples to the anharmonicity and causes the axial frequency shift.
Because of the slow synchrotron-radiation rate at low magnetic fields and the current high physical temperature of $T=4.5$~K, careful tuning of the cyclotron drive power is required to decouple the cyclotron radius and the magnetron radius.
This study should be possible with a dilution refrigerator and with a dedicated microwave drive line.

\section{Discussion}
\label{Sec:Discussion}
We first comment on our choice of $\rho_2$ and $z_0$.
For any value of $\rho_2$, one can choose $z_0$ to achieve the orthogonality condition ($D_2=0$ and $D_4\neq0$).
Indeed, one could choose $\rho_2/\rho_1=3.06$ and $z_0/\rho_1=2.26$ to achieve even better harmonicity  than in Tab.~\ref{tab:anharmonicity} with $C_6=0$.
However, doing so will increase $z_0$ to 4.52~mm and reduce the image charge pickup constant to $c_1=0.320$ [Eq.~\eqref{eq:gamma_z}], overall reducing $\gamma_z$ by a factor of 7 and making axial frequency determination more sensitive to voltage fluctuations and magnetron drift.
This concern can be solved by, for example, a stable liquid-helium-cooled cryocooler or a dilution refrigerator to further lower $T_z$.
Alternatively, one could reduce $z_0$ to increase $\gamma_z$.
Ultimately, how much one can reduce $z_0$ (and $\rho_1$ and $\rho_2$ correspondingly) will be limited by the uncontrolled electric charges on the insulation gaps and patch potentials, and it will be interesting to study this limit for fundamental precision measurements and for QIS.

The next step for the single-electron planar trap will be to observe the cyclotron and spin flip transitions.
This is possible using magnetic-bottle detection \cite{DehmeltMagneticBottle,RevModPhys.58.233}.
\begin{figure}[t]
    \centering
\includegraphics[width=\linewidth]{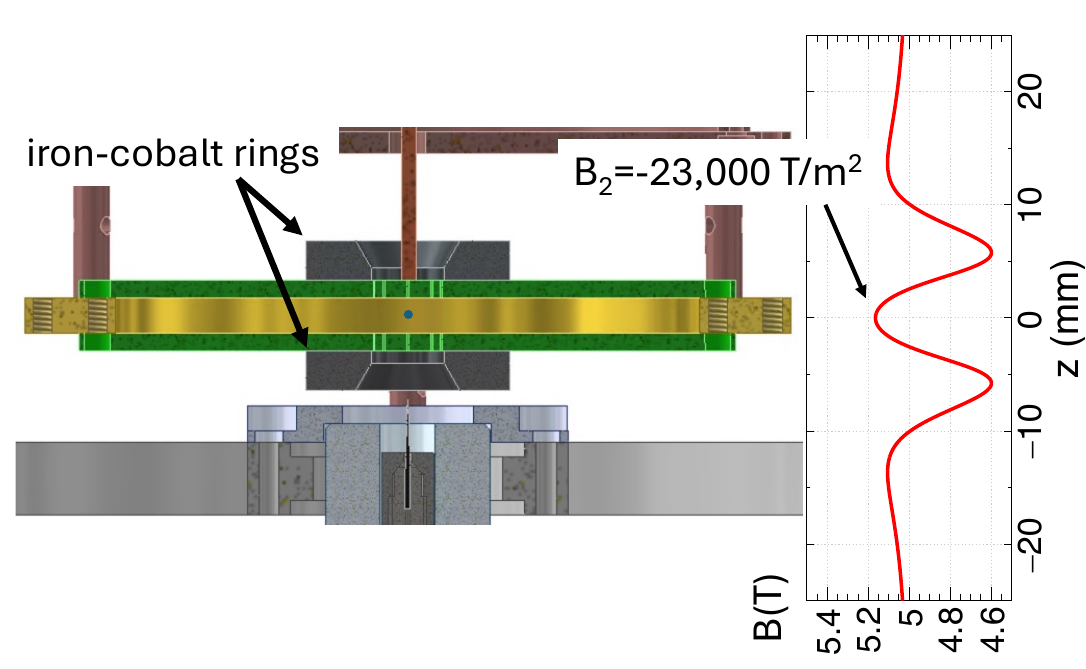} 
     \caption{Proposed magnetic bottle location in a $B=5$~T magnetic field. By placing iron-cobalt rings above and below the PCBs, one can achieve a large gradient $B_2=-23,000$~T/m$^2$.}
    \label{fig:bottle}
\end{figure}
One could integrate iron-cobalt rings on the top and bottom of the PCBs, creating a quadratic magnetic field gradient as large as $B_2=-23,000$~T/m$^2$ \cite{DehmeltMagneticBottle,peil1999observing,ulmer2011Thesis} (Fig.~\ref{fig:bottle}).
This will induce an axial frequency shift $\delta$ per cyclotron or spin flip transition,
\begin{equation}
    \frac{\delta}{2\pi}=\frac{1}{2\pi}\frac{\hbar eB_2}{m_e^2\omega_z}=-146~\text{Hz},
\end{equation}
which should be easily observable.
The magnitude of $B_2$ is proportional to $z_0^{-2}$, so an even faster readout could be possible with a smaller trap.
Currently, our trap is physically at $T=4.5$~K, making quantum cyclotron and spin-flip detection difficult, but this can be avoided by using a dilution refrigerator \cite{peil1999observing}.

Another important milestone will be the coupling of remote electrons mediated by image charges \cite{sorensen2004capacitive,QLS_Electron_2025}.
For the planar geometry, the image-charge-coupling rate is given by \cite{QLS_Electron_2025,QLS_PhysRevA_1990}
\begin{equation}
\omega_\textrm{ex}=\frac{e^2c_1^2}{8m_ez_0^2}|\textrm{Im}\left[Z(\omega_z)\right]|,\label{eq:exchangeRate}
\end{equation}
where $Z(\omega_z)$ is the impedance between the wired electrodes and ground.
The impedance can be controlled using a high-$Q$ resonator and by detuning $\omega_z$ from its resonance.
Again, a smaller $z_0$ will be helpful for achieving a large coupling rate.

\begin{figure}[t]
    \centering
\includegraphics[width=0.8\linewidth]{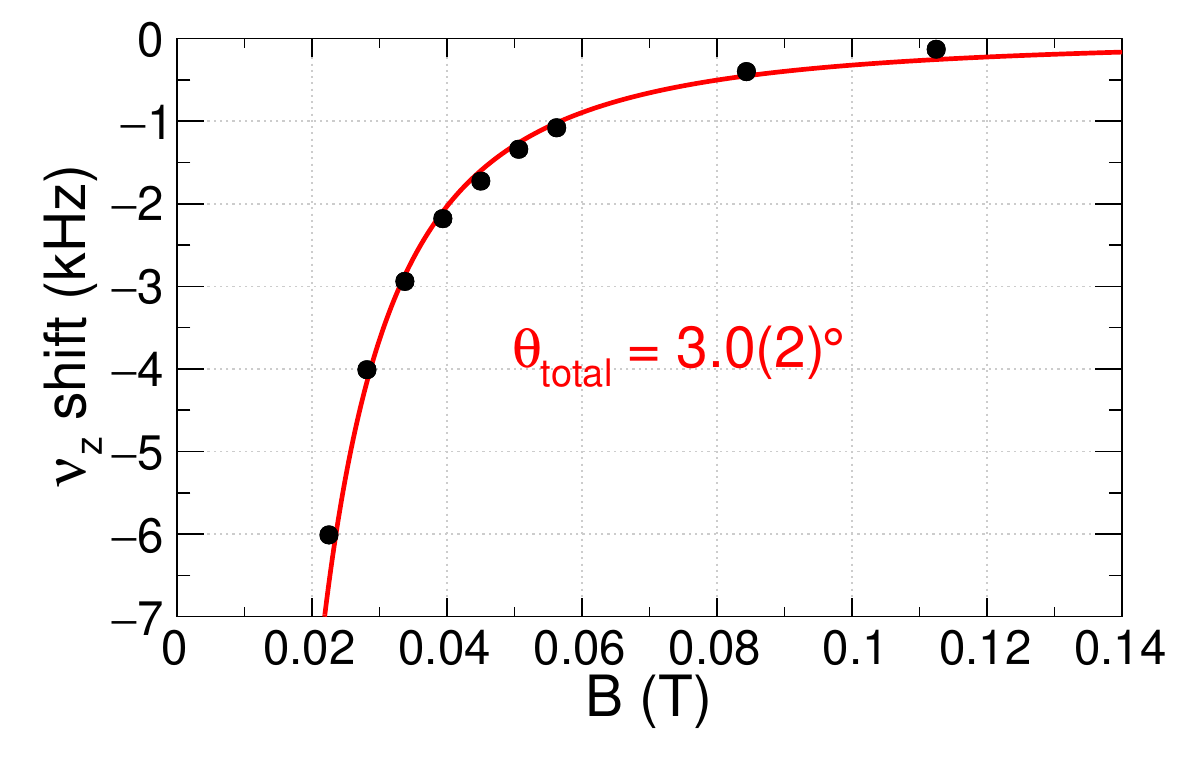} 
     \caption{Determination of the misalignment between the trap's electrostatic axis and the magnetic field axis using the axial frequency shift at low magnetic fields.}
    \label{fig:B-dependent-shift}
\end{figure}
For the dual-plane geometry, compared to a single-plane geometry, there could be a concern about the alignment of the two planes.
We argue that the leading-order effect from the misalignment between two planes can be compensated by adjusting the direction of the magnetic field using a vector magnet \cite{bluefors_vector_magnets,daprato2025vector}.
In our current design, the alignment of the two planes is simply defined by the through hole diameters for \#2-56 imperial screws (0.10-mm radial clearance), which could cause a misalignment of $\theta_\text{planes}=0.97$~degree.
The total misalignment $\theta_\text{total}$ is the sum of $\theta_\text{planes}$ and the alignment between the trap stack and the solenoid magnetic field.
$\theta_\text{total}$ causes $B$-field-dependent $\omega_z$ shift \cite{brown1982Gabrielse_InvarianceTheorem} and we are able to measure this at low magnetic fields to determine $\theta_\text{total}=3.0(2)$~degree (Fig.~\ref{fig:B-dependent-shift}).
The measured $\theta_\text{total}$ is larger than the possible misalignment of the two planes, presumably caused by the mounting of the trap stack with respect to the superconducting solenoid's axis.
If this becomes a concern, one can use a set of shim coils \cite{romeo1984magnet,fan2019gaseous,van1999ultrastable} or a standard vector magnet \cite{bluefors_vector_magnets} and perform a similar measurement to correct the effect of the dual-plane misalignment.

The collision-induced growth of the magnetron radius could be a problem for QIS using coupled magnetron motions at low magnetic fields \cite{lamata2010towards}.
Within our measurement accuracy, we did not observe this effect for $N=1$.
The lowest magnetic field we tried was $B=0.011$~T.
At this magnetic field, the cyclotron frequency is as low as $\omega_c/(2\pi)=300$~MHz, and its synchrotron radiation cooling becomes very slow, preventing us from stably measuring the axial frequency.
Since the single-electron behavior can be described analytically \cite{RevModPhys.58.233}, collision-induced magnetron-radius growth is not expected to be problematic except at extremely low magnetic fields or in multi-electron operation.

\section{Conclusion}
\label{Sec:Summary}
In conclusion, we have demonstrated single-electron trapping and detection in a dual-plane PCB Penning trap.
The mirror-symmetric geometry provides a highly harmonic and controllable electrostatic potential, enabling deterministic electron loading, measurement of the axial damping rate, detection of a single-electron axial dip, and measurement of the axial temperature.
We have also observed collision-induced growth of the magnetron radius at low magnetic fields, identifying an important consideration for future multi-electron operation.
These results are the first steps toward a scalable planar Penning-trap platform and provide a path toward cyclotron and spin-state detection, image-charge-mediated coupling, and quantum information science with trapped electrons.

\begin{acknowledgments}
We thank G. Gabrielse, J. Home, K. Taniguchi, A. Huang, B. Yu, I. Sacksteder, and H. H{\"a}ffner for fruitful discussions.
This work is supported by the National Science Foundation Award No.~2317134. X.F. acknowledges support from the Masason Foundation.
\end{acknowledgments}

\bibliography{PenningTrapExperimentRefs}

\end{document}